\documentclass[english,twocolumn]{article}
\usepackage[utf8]{inputenc}
\usepackage[big,online]{biosignale}
\usepackage[sort&compress,square,numbers]{natbib}
\usepackage{times}

\begin{document}
\papername{}
\paperyear{2020}

%--------------------------------------------------------------------------------------------------
% Bitte fügen Sie ab hier Ihr Manuskript ein
%--------------------------------------------------------------------------------------------------
\title{Reconstruction of Potassium Concentrations with the ECG on Imbalanced Datasets}
\runningtitle{Reconstruction of Potassium Concentrations with the ECG}
%\subtitle{The Influence of Dataset Imhomogenity on the Results}

\author*[1]{Nicolas Pilia}
\author[3]{Cristiana Corsi} 
\author[3]{Stefano Severi} 
\author[2]{Olaf Dössel} 
\author[2]{Axel Loewe}
\runningauthor{N. Pilia et al.}

\affil[1]{\protect\raggedright 
  Karlsruhe Institute of Technology (KIT), Institute of Biomedical Engineering, Karlsruhe, Germany. publications@ibt.kit.edu}
\affil[2]{\protect\raggedright
  Karlsruhe Institute of Technology (KIT), Institute of Biomedical Engineering, Karlsruhe, Germany.}
  \affil[3]{\protect\raggedright
  University of Bologna, Department of Electrical, Electronic, and Information Engineering ``Guglielmo Marconi'', Cesena, Italy.}
	
	
\transabstract{}
\vspace{-1em}
\maketitle

\section{Introduction} 
In Europe, the prevalence of chronic kidney disease (CKD) is estimated to be approximately 18.38\% [1], so it is one of the major challenges for health care systems \citep{a1}. In the end-stage, patients are facing a 30\% rise for the risk of lethal cardiac events (LCE) compared to non-CKD patients \citep{a2}. At the same time, end-stage CKD patients undergoing dialysis experience shifts in the potassium concentrations (e.g. in \citep{a3} [K$^+$]$_\text{o}$ pre-dialysis 5.23$\pm$0.76 mmol/l, post-dialysis 3.67$\pm$0.25 mmol/l). The increased risk of LCE paired with the changing ionic concentrations suggest a connection between LCE and concentration disbalances. However, this link has to be proven. To achieve this, a continuous monitoring device for the ionic concentrations is needed.\\
It is known that [K$^+$]$_\text{o}$ changes also manifest in the electrocardiogram (ECG). In the past, groups used these electrographic changes captured with one feature and reconstructed [K$^+$]$_\text{o}$ with a mean absolute error of 0.46$\pm$0.39\,mmol/l \citep{a4}. The question arises, if an optimised signal processing chain developed in previous works \citep{a5} can improve the result. Furthermore, we want to quantify the influence of a disbalanced training dataset on the final estimation result. The findings of this study can contribute to the overall goal to better predict and prevent LCE in CKD patients.

\vspace{-0.5em}
%%%%%%%%%%%%%%
\section{Methods}
\vspace{-0.25em}
\subsection{Dataset}
\vspace{-0.5em}
The study was performed on a dataset consisting of 12-lead ECGs recorded during dialysis sessions of 32 patients (two to three sessions per patient). The records therefore intrinsically include varying ionic concentrations (mainly [K$^+$]$_\text{o}$) due to the dialysis process. The concentrations were measured five to six times in each session by analysing blood samples at distinct time points.
\vspace{-0.25em}
\subsection{Preprocessing and Feature Extraction}
\begin{figure}
	\includegraphics[width=1\columnwidth]{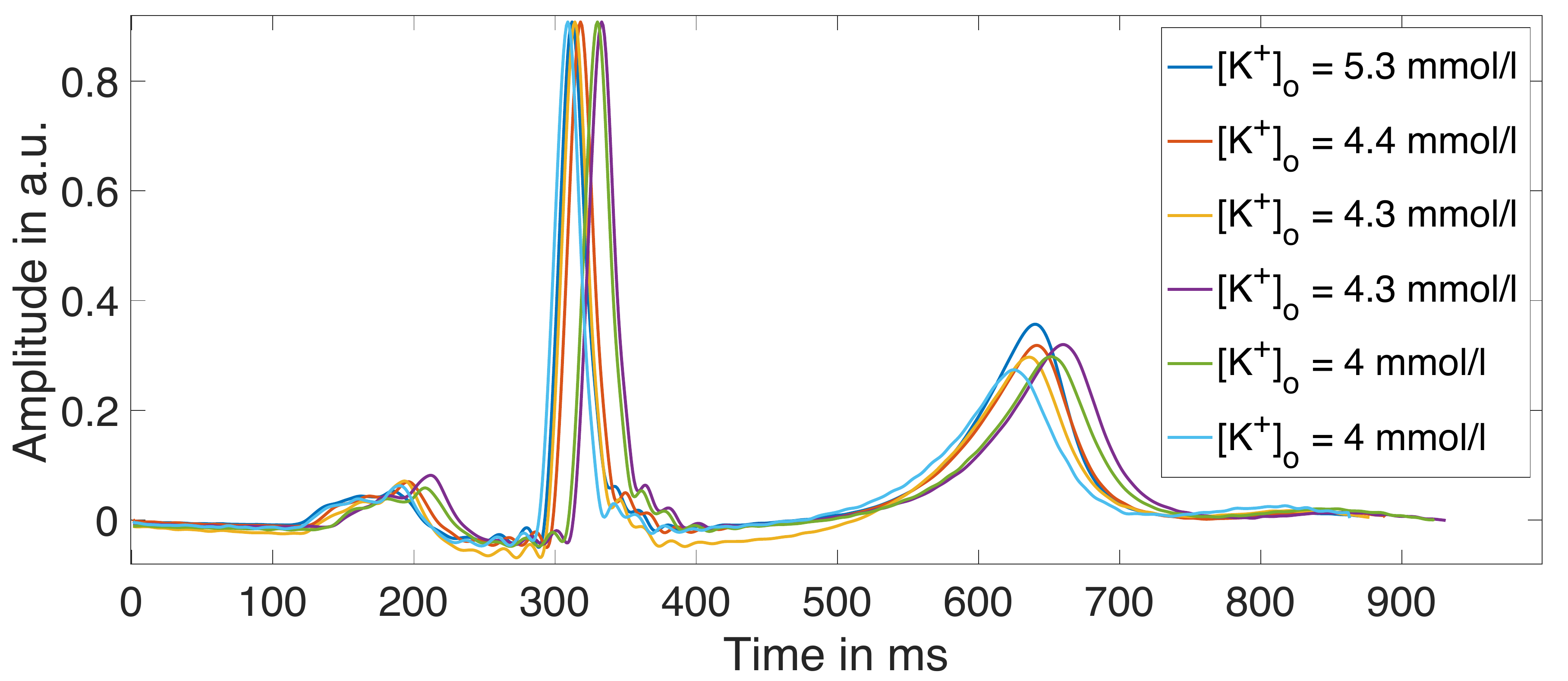}
	\caption{Templates of segments with different [K$^+$]$_\text{o}$ in one dialysis session.}
	\label{img:Templates}
	\vspace{-1em}
\end{figure}

\vspace{-0.5em}
After applying a Butterworth highpass filter (0.3\,Hz), a Butterworth lowpass filter (80\,Hz) and a spectral Gaussian notch filter (50\,Hz, standard deviation 1\,Hz), the 8 independent ECG-lead signals were split into segments around every measuring time instant each with a total segment length of 4\,min. These segments were used to build a beat template for every concentration measurement. We ended up with a dataset of beat templates and the respective measured ionic concentration. As we had 8 leads recorded, we further reduced the number of templates by transforming the 8 leads into one showing maximum T wave amplitude (as described in \citep{a5}, see also figure \ref{img:Templates}). After the lead reduction, we analysed every beat template based on the features proposed in \citep{a5}.

\vspace{-0.25em}
\subsection{Reconstruction}
\begin{figure}
	\includegraphics[width=1\columnwidth]{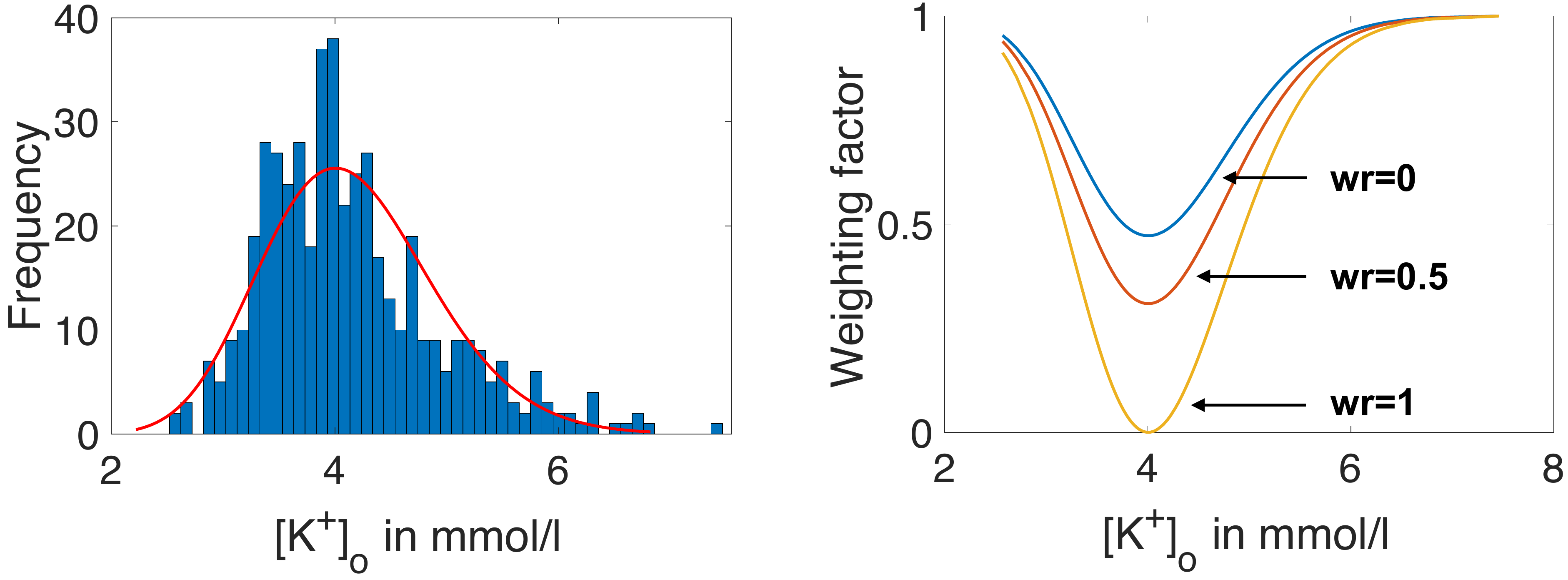}
	\caption{Left: Distribution of concentration values in the dataset; Right: respective weighting for every concentration for different weighting ratios $wr$.}
	\label{img:Weight}
	\vspace{-1em}
\end{figure}

\vspace{-0.5em}
We selected three features to find a mapping from ECG features to [K$^+$]$_\text{o}$: T-wave ascending slope, T-wave descending slope and T-wave amplitude. A polynomial model of 3rd order was used to reconstruct the concentrations from these features. To find the model, we solved a regularised weighted least squares problem:
\begin{equation}
    \min_x ||W \cdot (Ax-b)||_2 + \lambda ||x||_1
\end{equation}
$A$ is the Vandermonde matrix containing the feature values, $b$ is the vector of the ionic concentrations, $x$ is the model coefficient vector and $W$ is the error weighting vector. $\lambda$ was empirically chosen as 0.9.\\
The weighting matrix was determined by using the frequency of each concentration value in the dataset, weighting frequent concentration values less and rare concentrations more. By doing so, we tried to generate a model being suitable for the whole range of the concentrations and not only the normal concentrations that were very frequent in the dataset (see figure \ref{img:Weight} left). The weighting ratio $wr$ that scales the weighing curve for the different concentrations in the dataset was varied during the study, too. $wr=1$ means that the most frequent concentration value has a weighting factor of 0, $wr=0$ sets the weighting to approximately 0.5 (see figure \ref{img:Weight} right).

For every patient, we calculated a patient specific offset: We therefore applied the model found with all the other patients to this first session and calculated the concentration error. This was then applied to correct the estimation result of all other sessions of the specific patient. Only sessions other than the first session were used during cross-validation.

%%%%%%%%%%%%%%
\vspace{-0.5em}
\section{Results}

The estimation results are given in table \ref{tab:ResultsError} for different weighting vectors (weighting ratios $wr$) for the least squares problem. Here, we distinguish between a general result over all validation data and those lower or higher than 5\,mmol/l. With increasing $wr$, errors are increasing for the whole dataset. For the data partition with [K$^+$]$_\text{o}$ $<$5\,mmol/l, errors are increasing, for [K$^+$]$_\text{o}\geq$5\,mmol/l, errors are decreasing. Taking the histogram (figure \ref{img:Weight}) into account, [K$^+$]$_\text{o}$ $<$5\,mmol/l is more frequent in the dataset.

\begin{table}
\caption{Mean absolute estimation errors and their standard deviation (std) for different least squares weighting matrices.}
\begin{tabular}{l|ccc}
Weight & mean$\pm$std & mean$\pm$std &  mean$\pm$std 	\\ 
 matrix & $<$5\,mmol/l & $\geq$ 5\,mmol/l	&  all		\\ \midrule
no weights & 0.35$\pm$0.26 & 0.87$\pm$0.55 & 0.43$\pm$0.37\\
wr=0 & 0.34$\pm$0.26 &	0.80$\pm$0.52 &0.41$\pm$0.35\\
wr=0.5 & 0.37$\pm$0.28 &	0.79$\pm$0.50 & 0.43$\pm$0.35\\
wr=1  & 0.46$\pm$0.36 &	0.74$\pm$0.47 & 0.50$\pm$0.39\\
\end{tabular}
\label{tab:ResultsError}
\vspace{-0.5em}
\end{table}

%no weights & 0.3515$\pm$0.2628 & 0.8715$\pm$0.5461 & 0.4278$\pm$0.3685\\
%wr=0 & 0.3436$\pm$0.2582 &	0.8010$\pm$0.5249 &0.4107$\pm$0.3504\\
%wr=0.5 & 0.3692$\pm$0.2826 &	0.7866$\pm$0.4961 & 0.4304$\pm$0.3541\\
%wr=1  & 0.4616$\pm$0.3617 &	0.7406$\pm$0.4677 & 0.5025$\pm$0.3909\\

%%%%%%%%%%%%%%
\vspace{-0.5em}
\section{Discussion and Outlook}
Compared to the results in \citep{a4}, the results for the whole dataset are in the same range regarding the concentration estimation errors. This comparison was made using a dataset including the data from \citep{a4}, nevertheless with additional data and other features. The calculated least squares errors given in table \ref{tab:ResultsError} were not weighted w.r.t. their frequency in the dataset to be still comparable to other works in contrast to the sum of least squares during model fitting. This partly explains the decreasing performance on the whole dataset with increasing $wr$. It could be interesting to compare weighted errors for different methods to estimate the generalisation capability of the different methods.\\
However, and apart from the exact reconstruction results, we can conclude that a model being valid for all patients and not only the majority, needs to be learned with a more homogeneous dataset. This can be achieved by leaving out data points or by weighting the errors during the model fitting. Table \ref{tab:ResultsError} reflects this. With increasing weighting, we increase the performance on the part of the [K$^+$]$_\text{o}$ that are less frequent. In the end, $wr$ has to be chosen according to the desired application. In our case, we want to detect unphysiologically high [K$^+$]$_\text{o}$, so we need a model being able to detect those [K$^+$]$_\text{o}$.\\
In a next step, we need to include the weighting procedure in other reconstruction techniques, such as neural networks as used in \citep{a5}.

\end{document}